\documentclass{amsart}

\usepackage{amsmath}
\usepackage{amsthm}
\usepackage{amsfonts}
\usepackage{amssymb}
\usepackage{epsfig}
\usepackage{graphicx} 
\usepackage{palatino,url}

\theoremstyle{plain}

\theoremstyle{remark}

\theoremstyle{definition}

\include{psfig}

\begin{document}
\title{Inter Genre Similarity Modelling For Automatic Music Genre Classification}
%
\author[Ba\u{g}c\i]{Ula\c{s} Ba\u{g}c\i}
\address{Collaborative Medical Image Analysis on Grid (CMIAG),The University of Nottingham, Nottingham, UK}
\email{ulasbagci@ieee.org}
\urladdr{www.ulasbagci.net}

\author[Erzin]{Engin Erzin}
\address{MVGL,College of Engineering, Ko\c{c} University,Istanbul,Turkey}
\email{eerzin@ku.edu.tr}

\begin{abstract}
Music genre classification is an essential tool for music
information retrieval systems and it has been finding critical
applications in various media platforms. Two important problems of
the automatic music genre classification are feature extraction
and classifier design. This paper investigates inter-genre
similarity modelling (IGS) to improve the performance of automatic
music genre classification. Inter-genre similarity information is
extracted over the mis-classified feature population. Once the
inter-genre similarity is modelled, elimination of the inter-genre
similarity reduces the inter-genre confusion and improves the
identification rates. Inter-genre similarity modelling is further
improved with iterative IGS modelling(IIGS) and score modelling
for IGS elimination(SMIGS). Experimental results with promising
classification improvements are provided.
\end{abstract}

\maketitle
\section{Introduction}
Music genre classification is crucial for the categorization of
bulky amount of music content. Automatic music genre
classification finds important applications in professional media
production, radio stations, audio-visual archive management,
entertainment and recently appeared on the Internet. Although
music genre classification is done mainly by hand and it is hard
to precisely define the specific content of a music genre, it is
generally agreed that audio signals of music belonging to the same
genre contain certain common characteristics since they are
composed of similar types of instruments and having similar
rhythmic patterns. These common characteristics motivated recent
research activities to improve automatic music genre
classification \cite{paper:20:tzanetakis02,
paper:20:pye00:_conten,paper:20:li00:_compar,paper:20:lippens04}.
The problem is inherently challenging as the human identification
rates after listening to $3$sec samples are reported to be around
$70$\% \cite{paper:20:perrot99}.

Feature extraction and classifier design are two pretentious
problems of the automatic music genre classification. Timbral
texture features representing short-time spectral information,
rhythmic content features including beat and tempo, and  pitch
content features are investigated throughly in
\cite{paper:20:tzanetakis02}. Another novel feature extraction
method is proposed in \cite{paper:20:li00:_compar}, in which local
and global information of music signals are captured by
computation of histograms on their Daubechies wavelets
coefficients. A comparison of human and automatic music genre
classification is presented in a study \cite{paper:20:lippens04}.
Mel-frequency cepstral coefficients (MFCC) are also used for
modelling and discrimination of music signals
\cite{paper:20:tzanetakis02,paper:20:logan97:_mel}.

Various classifiers are employed for automatic music genre
recognition including K-Nearest Neighbor (KNN) and Gaussian
Mixture Models (GMM) classifiers as in
\cite{paper:20:tzanetakis02,paper:20:li00:_compar}, and Support
Vector Machines (SVM) as in \cite{paper:20:li00:_compar}. In a
recent study,  Boosting is used as a dimension reduction tool for
audio classification \cite{paper:20:ravindran04}.

In \cite{paper:20:ulas05}, we proposed \emph{Boosting Classifiers}
to improve the automatic music genre classification rates. In this
study, in addition to inter-genre similarity modelling, two
alternative classifier structures are proposed: i) Iterative
inter-genre similarity modelling, and ii) Score modelling for
inter-genre similarity elimination. Once the inter-genre
similarity is modelled, elimination of those similarities from the
decision process reduces the inter-genre confusion and improves
the identification rates.

The organization of the paper includes a brief description of the
feature extraction in Section \ref{sec:feat}. The discriminative
music genre classification using the inter-genre similarity is
discussed in Section~\ref{sec:DGC}. Later in Section~\ref{sec:exp}
experimental results are provided following with discussions and
conclusions in the last section.

\section{Feature Extraction}
\label{sec:feat}

Timbral texture features, which are similar to the proposed
feature representation in \cite{paper:20:tzanetakis02}, are
considered in this study to represent music genre types in the
spectral sense. Short-time analysis over 25ms overlapping audio
windows are performed for the extraction of timbral texture
features for each 10ms frame. Hamming window of size 25ms is
applied to the analysis audio segment to remove edge effects. The
resulting timbral features from the analysis window are combined
in a $17$ dimensional vector including the first $13$ MFCC
coefficients, zero-crossing rate, spectral centroid, spectral
roll-off and spectral flux.

\section{Music Classification using Inter-Genre Similarity}
\label{sec:DGC}

The music signals belonging to the same genre contain certain
common characteristics as they are composed of similar types of
instruments with similar rhythmic patterns. These common
characteristics are captured with statistical pattern recognition
methods to achieve the automatic music genre classification
\cite{paper:20:tzanetakis02,paper:20:pye00:_conten,paper:20:li00:_compar,paper:20:lippens04}.
The music genre classification is challenging problem, especially
when the decision window spans a short duration, such as a couple
of seconds. One can also expect to observe similarities of
spectral content and rhythmic patterns across different music
genre types, and with a short decision window mis-classification
and confusion rates increase. IGS modelling is proposed to
decrease the level of confusion across similar music genre types.
The IGS modelling forms clusters from the hard-to-classify samples
to further eliminate the inter-genre similarity in the decision
process of classification system. Inter-genre similarity modelling
is further improved with iterative IGS modelling(IIGS) and score
modelling for IGS elimination(SMIGS). IGS  and its variations are
described in the following subsections.

\subsection{Inter-Genre Similarity Modelling}
The timbral texture features represent the short-term spectral
content of music signals. Since, music signals may include similar
instruments and similar rhythmic patterns, no sharp boundaries
between certain different genre types exist. The inter-genre
similarity modelling (IGS) is proposed to capture the similar
spectral contents among different genre types. Once the IGS
clusters are statistically modelled with GMM, the IGS frames can
be captured and removed from the decision process to reduce the
inter-genre confusion.

Let $\lambda_1, \lambda_2,\dots,\lambda_N$ be the $N$ different
genre models in the database($N = 9$ in our database). In this
study, the Gaussian Mixture Models (GMM) are used for the
class-conditional probability density function estimation,
$p(f|\lambda)$, where $f$ and $\lambda$ are respectively the
feature vector and the genre class model.  The construction of the
IGS clusters and the class-conditional statistical modelling(GMM)
can be achieved with the following steps:
\begin{itemize}
\item [i.] Perform the statistical modelling  of each genre with
GMM in the database using the available training data of the
corresponding music genre class.
\item [ii.] Perform frame based
genre identification task over the training data {\it ,(gmm test
over training data)}, and label each frame as a
true-classification or mis-classification.
 \item [iii.] Construct the statistical model,
$\lambda_{IGS}$ with GMM, through the IGS cluster over all the
mis-classified frames among all the music genre types. \item [iv.]
Update all the $N$-class music genre models, $\lambda_n$, using
the true-classified frames.
\end{itemize}

The above construction creates $N$-class music genre models and a
single-class IGS model. In the music genre identification process,
given a sequence of features, $\{ f_1, f_2,\dots,f_K \} $, which
are extracted from a window of music signal, one can find the most
likely music genre class, $\lambda^*$, by maximizing the weighted
joint class-conditional probability,
\begin{equation}
\label{eqn:wdec} \lambda^* = \arg\max_{\lambda_n} \frac{1}{\sum_k
\omega_{kn}} \sum_{k=1}^{K} \omega_{kn} \log p(f_k | \lambda_n)
\end{equation}
where the weights $\omega_{kn}$ are defined based on the
class-conditional IGS model as following,
\begin{equation}
\omega_{kn} = \begin{cases}
1 & \mbox{if } p(f_k|\lambda_n) > p(f_k|\lambda_{IGS}),\\
0 & \mbox{otherwise}.
              \end{cases}
\end{equation}
The proposed weighted joint class-conditional probability
maximization eliminates the IGS frames for each music genre from
the decision process. The inter-genre confusion decreases and the
genre classification rate increases with the resulting
discriminative decision process. Experimental results, which are
supporting the discrimination based on the IGS elimination, are
presented in Section \ref{sec:exp}.

\subsection{Iterative Inter-Genre Similarity Modelling}
\label{ssec:iterative} Inter-genre similarity modelling can be
repeatedly used to extend the detection of hard-to-classify
samples in the training data. In each iteration a new IGS model is
formed over the new set of mis-classified samples. In the decision
process a frame is eliminated if it matches any one of the IGS
classes.  The construction of the $T$-step iterative inter-genre
similarity (IIGS) models can be defined with the following steps:
\begin{itemize}
\item [i.] Perform IGS modelling, and get $\lambda_{IGS_1}$ and
updated music genre models $\lambda_n$ for all $N$-class. Set
$t=2$. \item [ii.] Perform frame based genre identification task
with IGS model over the training data, and label each frame as a
true-classification, mis-classification or true-IGS
classification. \item [iii.] Construct the statistical model,
$\lambda_{IGS_t}$, over all the mis-classified frames among all
the music genre types. \item [iv.] Update all the $N$-class music
genre models, $\lambda_n$, over the true-classified frames.
\item[v.] Increment iteration counter $t$, and if $t \leq T$ go to
step [ii.]
\end{itemize}

The above construction creates $N$-class music genre models and
$T$-class IGS model. In the music genre identification process,
the decision is taken by maximizing the weighted joint
class-conditional probability in Equation~\ref{eqn:wdec}. In IIGS
modelling the weights $\omega_{kn}$ are re-defined based on the
IGS models as following,
\begin{equation}
\omega_{kn} = \begin{cases}
1 & \mbox{if } p(f_k|\lambda_n) > p(f_k|\lambda_{IGS_t}) \mbox{for any $t=1,...,T$}\\
0 & \mbox{otherwise}.
                \end{cases}
\end{equation}

\subsection{Score Modelling for IGS Elimination (SMIGS)}
\label{ssec:disc} In IGS modelling, the elimination of likely
mis-classified frames is performed with a hard thresholding. That
is, if IGS model produces the highest likelihood score for a
frame, that frame is eliminated even though the best genre model
is the true class and has a very close likelihood score to the IGS
likelihood. Alternatively, IGS may score lower than the best
likelihood score, while the best likelihood score belongs to a
false-class. For such possible wrong decisions, the relative
likelihood differences can be used as a reliability factor for the
IGS elimination process. For example, if the likelihood difference
between best two likelihood scores of music genre classes is low,
one can claim that the decision is not reliable and the frame
should be eliminated even though the IGS likelihood score is not
the highest one. Hence, rather than taking a hard decision based
on IGS likelihood score, one can model frame elimination based on
the likelihood score distributions of the best $M$ genre classes
and the IGS class. Score modelling for IGS elimination is built on
this idea, and described with the following procedure:
\begin{itemize}
\item[i.] Perform IGS modelling, and get $\lambda_{IGS}$ and
updated music genre models $\lambda_n$ for all $N$-class.
\item[ii.] Perform frame based genre identification task for each genre $\lambda_n$ over the training data, extract the highest class conditional likelihood score that belongs to a class $\lambda^*$ other than $\lambda_n$, $p(f_k|\lambda^*)$, $p(f_k|\lambda_n)$ and $p(f_k|\lambda_{IGS})$. Form a 3-dimensional likelihood score difference vector, \\
\[ s_k = [ \Delta_k^1 \; \Delta_k^2 \; \Delta_k^3 ] \]\\
where
\begin{eqnarray*}
\Delta_k^1 &=& (p(f_k|\lambda_n)-p(f_k|\lambda^*)), \\
\Delta_k^2 &=& (p(f_k|\lambda_n)-p(f_k|\lambda_{IGS})), \\
\Delta_k^3 &=& (p(f_k|\lambda^*)-p(f_k|\lambda_{IGS})).
\end{eqnarray*}
\item[iii.] Construct the statistical models of the likelihood
difference vectors for each genre $\lambda_n$ over the
true-classified and mis-classified frames of IGS modelling in step
[i.], respectively as $\lambda_{n_1}$ and $\lambda_{n_0}$.
\end{itemize}

The above construction creates $N$-class music genre models and
for each genre true- and mis- classification models of the
likelihood differences are extracted. In the music genre
identification process, the decision is taken by maximizing the
weighted joint class-conditional probability in
Equation~\ref{eqn:wdec}. In score modelling for IGS elimination
the weights $\omega_{kn}$ are re-defined as following,
\begin{equation}
\omega_{kn} = \begin{cases}
1 & \mbox{if } p(f_k|\lambda_{n_1}) > p(f_k|\lambda_{n_0})\\
0 & \mbox{otherwise}.
              \end{cases}
\end{equation}

\section{EXPERIMENTAL RESULTS}
\label{sec:exp} Evaluation of the proposed classification
algorithms is performed over a music genre database that includes
$9$ different genre types: classical, country,  disco, hip-hop,
jazz, metal, pop, reggae and rock. The songs in the database had
been collected from CD collection of the authors and most of the
songs are recorded from randomly chosen broadcast radios on the
internet. The database includes totally $566$ different
representative audio segments of duration $30$sec for all $9$
music genre types, resulting a total duration of $566 \times 30 =
16980$ seconds. All the audio files are stored mono at 16000Hz
with $16$-bit words. The resulting timbral texture feature vectors
are extracted for each $25$msec audio frame with an overlapping
window of size $10$msec. The music genre classification is
performed based on the maximization of the class-conditional
probability density functions, which are modelled using the
Gaussian mixture models (GMM). In experiments two-fold cross
validation is used: the database is split into two partitions,
each partition includes half of the audio segments from each genre
type, where they are used in alternating order for training and
testing of the music genre classifiers, and the average correct
classification rates are reported in the following.

\begin{table}[htdp]
\begin{center}
\begin{tabular}{||c|c|c|c|c||}  \hline\hline
Decision & \multicolumn{4}{|c|}{Correct Classification Rates (\%)}\\
\hline Window  & Flat     &  IGS   & SMIGS  & IIGS      \\ \hline

0.5s    & 44.61 &  49.95 &  52.70    &  50.46        \\ \hline 1s
& 46.74 &  54.08 &  55.62    &  54.35         \\ \hline 3s  &
49.22 & 58.50 &  61.99    &  59.16        \\ \hline 30s & 55.73 &
62.56 & 62.57    &  64.71        \\ \hline\hline
\end{tabular}
\caption{{\it The average correct classification rates of the
flat, IGS clustering, score modeling for IGS elimination (SMIGS)
and iterative IGS clustering, using 8-mixture GMM modeling for
varying decision window sizes.} \label{tbl:gmm8}}
\end{center}
\end{table}
The three proposed music genre classification schemes, which are
variations of inter-genre similarity elimination to reduce
inter-genre confusion, are evaluated and compared with a flat
classifier structure. Tables~\ref{tbl:gmm8}, \ref{tbl:gmm16} and
\ref{tbl:gmm32} present correct classification rates of flat, IGS,
SMIGS and IIGS classifiers for respectively 8, 16 and 32 mixture
GMM modelling. Note that, IGS classification improves flat
classification rates for all decision windows. We also observe
further improvement using IIGS and SMIGS classifiers over IGS
classifier. While the IIGS improvements are observed to be better
for larger decision window sizes, the SMIGS improvements are
better for smaller decision window sizes with 8-mixture GMM
modeling. This behavior is expected, since iterative IGS expands
the inter-genre similarity modelling, which causes elimination of
more frame decisions, IIGS needs larger decision window sizes to
bring further improvement.

\begin{table}[htdp]
\begin{center}
\begin{tabular}{||c|c|c|c|c||}  \hline\hline
Decision & \multicolumn{4}{|c|}{Correct Classification Rates (\%)}\\
\hline Window  &   Flat      & IGS  & SMIGS   & IIGS    \\ \hline

0.5s    & 46.87    &  53.49  & 55.35   & 55.09        \\ \hline 1s
& 49.32    &  56.80  & 57.06   & 58.71        \\ \hline 3s  &
53.18 &  61.89  & 61.53   & 64.23        \\ \hline 30s & 57.78
&  66.91 & 67.26   & 72.48        \\ \hline\hline
\end{tabular}
\caption{{\it The average correct classification rates of the
flat, IGS clustering, score modelling for IGS elimination (SMIGS)
and iterative IGS clustering, using 16-mixture GMM modelling for
varying decision window sizes.} \label{tbl:gmm16}}
\end{center}
\end{table}

However, with increasing number of GMM mixtures as presented in
Tables~\ref{tbl:gmm16} and \ref{tbl:gmm32}, iterative IGS
classifier is observed as the clear winner of these three
discriminative music genre classification schemes at all decision
window sizes. The classification rate improvements, which are
achieved with IGS based classifiers, are significant, especially
when the challenging automatic music genre classification task is
considered with 70\% human identification rate over 3s decision
windows and 61\% identification rate reported in
\cite{paper:20:tzanetakis02}.

\begin{table}[htdp]
\begin{center}
\begin{tabular}{||c|c|c|c|c||}  \hline\hline
Decision & \multicolumn{4}{|c|}{Correct Classification Rates (\%)}\\
\hline Window  &  Flat       & IGS   & SMIGS & IIGS     \\ \hline

0.5s    & 48.83        & 61.05      &  62.03  &  62.57
\\ \hline 1s  & 51.81        & 65.28      &  66.64  &  66.96
\\ \hline 3s  & 54.83        & 73.25      &  71.00  &  74.43
\\ \hline 30s & 59.90        & 83.16      &  81.58  &  84.41
\\ \hline\hline
\end{tabular}
\caption{{\it The average correct classification rates of the
flat, IGS clustering, score modelling for IGS elimination (SMIGS)
and iterative IGS clustering, using 32-mixture GMM modelling for
varying decision window sizes.} \label{tbl:gmm32}}
\end{center}
\end{table}

\section{Conclusion}
\label{sec:conc} Automatic music genre classification is an
important tool for music information retrieval systems. Feature
extraction and the classification design are the two significant
problem of music genre classification systems. In feature
extraction, a set of widely used timbral features are considered.
In this work, we investigate three novel classifier structures for
discriminative music genre classification. In proposed classifier
structure, inter-genre similarities are captured and modelled over
the mis-classified feature population for the elimination of the
inter-genre confusion. The proposed iterative IGS model expands
the inter-genre similarity modelling to better eliminate these
similarities for the decision process. In score modelling for IGS
elimination, a novel scheme based on statistical modelling of
decision region of each genre for capturing IGS frames is
presented. Experimental results with promising identification
improvements are obtained with classifier design based on the
similarity measure among genres. Both in \cite{paper:20:ulas05}
and in this study we observed that IGS improves the flat
classifiers and we also investigated the two extension of IGS
modelling which increments the identification rates further
dep

\section{Acknowledgements}
Ulas Bagci was with the College of Engineering, Koc University, Istanbul, 34450, Turkey, when the project was published.


\end{document}